\batchmode
\makeatletter
\def\input@path{{C:/Users/hp/Downloads/PRR/}}
\makeatother
\documentclass[letterpaper,twocolumn,reprint,amsmath,amssymb,prl]{revtex4-2}
\usepackage{mathptmx}

\usepackage[T1]{fontenc}
\usepackage[latin9]{inputenc}
\usepackage{color}
\usepackage{amsmath}
\usepackage{graphicx}
\usepackage{wasysym}
\usepackage[pdftex,unicode=true,pdfusetitle,
 bookmarks=true,bookmarksnumbered=false,bookmarksopen=false,
 breaklinks=false,pdfborder={0 0 0},pdfborderstyle={},backref=false,colorlinks=true]
 {hyperref}

\makeatletter

\pdfpageheight\paperheight
\pdfpagewidth\paperwidth

\makeatother

\begin{document}
\title{Atomic excitation delocalization at the clean to disordered interface
in a chirally-coupled atomic array}

\author{C.-C. Wu$,^{1}$}
\email{freddywu0811@gmail.com}
\author{ K.-T. Lin$,^{2}$ I G. N. Y. Handayana$,^{1,3,4}$
C.-H. Chien$,^{1,2}$ S. Goswami$,^{1}$ G.-D. Lin$,^{2,5,6}$Y.-C.
Chen,$^{1}$and H.-H Jen$^{1,5}$}
\email{sappyjen@gmail.com}
\affiliation{$^{1}$Institute of Atomic and Molecular Sciences, Academia Sinica,
Taipei 10617, Taiwan~\linebreak{}
$^{2}$Department of Physics and Center for Quantum Science and Engineering,
National Taiwan University, Taipei 10617, Taiwan~\linebreak{}
$^{3}$Molecular Science and Technology Program, Taiwan International
Graduate Program, Academia Sinica, Taiwan~\linebreak{}
$^{4}$Department of Physics, National Central University, Taoyuan
City 320317, Taiwan~\linebreak{}
$^{5}$Physics Division, National Center for Theoretical Sciences,
Taipei 10617, Taiwan~\linebreak{}
$^{6}$Trapped-Ion Quantum Computing Laboratory, Hon Hai Research
Institute, Taipei 11492, Taiwan}
\begin{abstract}
In one-dimensional quantum emitter systems, the dynamics of atomic
excitations are influenced by the collective coupling between emitters
through photon-mediated dipole-dipole interactions. By introducing
positional disorders in a portion of the atomic array, we investigate
the delocalization phenomena at the interface between disordered zone
and clean zone. The excitation is initialized as symmetric Dicke states
in the disordered zone, and several measures are used to quantify
the excitation localization. We first use population imbalance and
half-chain entropy to investigate the excitation dynamics under time
evolutions, and further investigate the crossover of excitation localization
to delocalization via the gap ratio from the eigenspectrum in the
reciprocal coupling case. In particular, we study the participation
ratio of the whole chain and the photon loss ratio between both ends
of the atomic chain, which can be used to quantify the delocalization
crossover in the non-reciprocal coupling cases. Furthermore, by increasing
the overall size or the ratio of the disordered zone under a fixed
number of the whole chain, we observe that excitation localization
occurs at a smaller disorder strength in the former case, while in
the latter, a facilitation of the delocalization appears when a significant
ratio of clean zone to disordered zone is applied. Our results can
reveal the competition between the clean zone and the disordered zone
sizes on localization phenomenon, give insights to non-equilibrium
dynamics in the emitter-waveguide interface, and provide potential
applications in quantum information processing.
\end{abstract}
\maketitle
\hypersetup{
colorlinks=true,
urlcolor=blue,
linkcolor=blue,
citecolor=blue
}

\section{Introduction\label{sec:level1}}

Quantum particle dynamics in the random potential of disordered lattices
have been extensively studied since Anderson's work on single particle
localization \citep{Anderson_1958}. In Anderson's primary work, even
though the spin can transport between lattice sites through dipolar
interaction, the spin diffusion remains absent when the random energy
is introduced from site to site. Besides Anderson localization in
a non-interacting regime \citep{Jendrzejewski_2012,Tsoi_2008,Kondov_2011,Lahini_2008,Schwartz_2007},
this single particle localization is still signified in an extensive
category of closed quantum system even when significant interactions
are involved \citep{Evers_2008,Cl_ment_2005,Schmitteckert_1998,Paul_2009}.
With these unique phenomena, abundant investigations have been explored
accordingly, like many-body localization \citep{Sierant_2022,Bar_Lev_2014}
and ergodicity in a closed system \citep{Serbyn_2021,Eckmann_1985,Turner_2018}.
Among this research, one of topics that has attracted much interest
in quantum dynamics is disordered-induced localization influenced
by additional baths \citep{Sels_2022,Luitz_2017,L_onard_2023,Morningstar_2022,De_Roeck_2017,Thiery_2018}.
Unlike fully disordered systems, once an additional bath interacts
with an initially localized system, the localization may break down
under the influence of the baths \citep{Sels_2022,Luitz_2017,Morningstar_2022,Thiery_2018,De_Roeck_2017}.
This has led to recent studies on quantum avalanches influenced by
the clean system size \citep{L_onard_2023,Thiery_2018}. Quantum avalanches
represent an accelerated bath penetration into the localization zone
as the clean system size increases. In the recent experiment \citep{L_onard_2023},
the localization is still robust under the influence of a small-size
bath, whereas the localization melts down acceleratedly when coupled
to a large-size bath.

Despite rich explorations in localization have been achieved, current
research is predominantly on finite range coupling and closed systems.
The localization behavior on an intrinsic open system is still less
investigated because the inevitable dissipation make the phase transition
from localization to delocalization hard to be identified. Until recently,
single particle localization and many-body localization have been
explored in an open system \citep{Fayard_2021,Jen_2020,L_schen_2017}.
However, whether an additional bath induces the localization breakdown
has also been scarcely explored in an open system, and the influence
of an additional bath is still unclear. In this study, we focus on
single excitation diffusion in a one-dimensional two-level quantum
emitters (TLQE) array coupled to the photonic waveguides \citep{Roy_2017,Sheremet_2023},
via the evanescent waves \citep{Bliokh_2015}. In addition to study
localization and non-equilibrium dynamics in a closed and finite range
coupling system \citep{Manasi_2018,See_2019}, a TLQE array also offers
an opportunity to investigate localization in an open and all-to-all
interaction system \citep{Hood_2016,Solano_2017,Tiranov_2023,Defenu_2023,Roy_2017,Sheremet_2023}.
Moreover, many related phenomena have already been explored in one-dimensional
equidistant TLQE with chiral couplings, like localization \citep{Jen_2020,Mahmoodian_2020,Jen_2021,Mirza_2017,Fayard_2021,Jen_2022},
photon transport \citep{Mahmoodian_2018,Liao_2015,Song_2018,Mahmoodian_2020,Guti_rrez_J_uregui_2022,Tsoi_2008,Masson_2020,Jen_2021},
photon storage \citep{Jen_2016,Jen_2017,Needham_2019,Jen2_2021},
excitation dynamics \citep{Jen_2020,Needham_2019,Douglas_2015,Henriet_2019,Jen_2019},
and photon-mediated localization \citep{Zhong_2020}. 

In this work, we apply disorders only to a half-chain of the TLQE
with chiral couplings \citep{Lodahl_2017}, creating an interface
between the disordered zone and the clean zone. We then investigate
the excitation distribution dynamics at long time and determine the
crossover of delocalization to disorder-induced localization phases
under the influence of clean zones. Additionally, we study photon
loss from the TLQE, which serves as a measure of delocalization with
initialized symmetric Dicke states in the disordered zone. Furthermore,
we examine the influences of system sizes on the phase boundary and
on the excitation distributions near the interface. We find that localization
is enhanced with a larger ratio of the disordered zone or with a larger
overall system size, while increasing the ratio of the clean zone
can further delocalize the excitations in the disordered zone when
a significant clean zone ratio is applied. Our work explores the modified
excitation localization at the clean-disordered interface in a chirally-coupled
atomic array, which can provide insights to non-equilibrium quantum
dynamics and controlled retention of initialized quantum states in
open quantum systems.

The paper is organized as follows. In Sec. \ref{sec:Theoretical-Model},
we introduce the theoretical model for a chirally-coupled one-dimensional
atomic chain and describe the system setup. In Sec. \ref{sec:local},
we compare the clean-disordered system with the fully disordered system
by population imbalance and half-chain entropy, and identify the localization
in the disordered zone and near the interface. In Sec. \ref{sec:Participation-Ratio},
we investigate the gap ratios and the participation ratio, both of
which can serve as measures for the excitation localization. In Sec.
\ref{sec:Directional-Photon-loss}, we investigate the directional
photon loss ratio, providing insights into the excitation dynamics
and serving as a measure of the localization crossover. In Sec. \ref{sec:Size-effects},
we further discuss the influence of the size of the clean and disordered
zones on the localization crossover point. In Sec. \ref{sec:Conclusion},
we summarize the localization properties in the TLQE system and explore
the feasibility of photon measurement in different platforms. We also
discuss potential future works with similar setups.

\section{Theoretical Model\label{sec:Theoretical-Model}}

\begin{figure*}[t]
\begin{centering}
\includegraphics[scale=0.3]{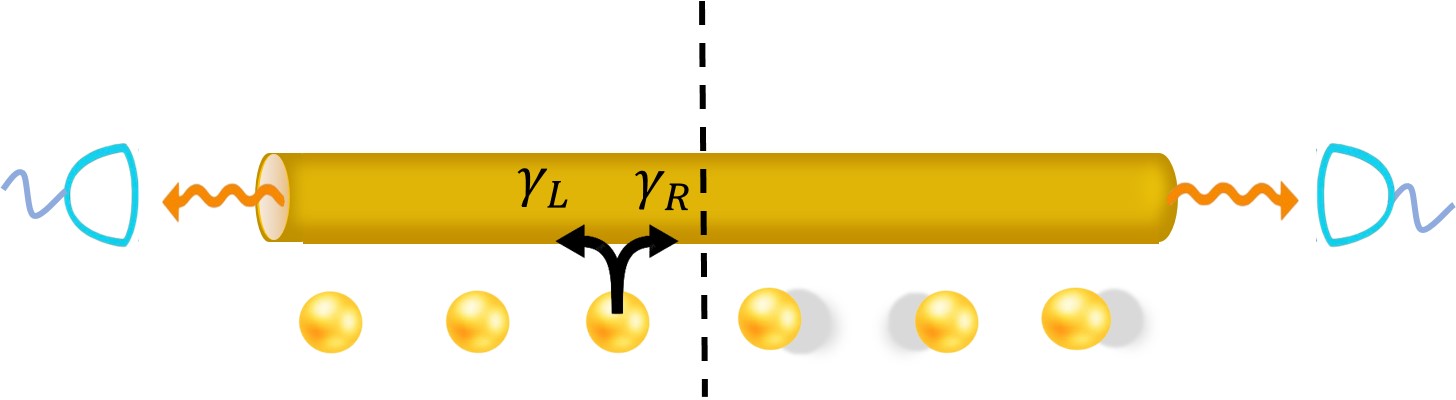}
\par\end{centering}
\begin{centering}
\includegraphics[scale=0.3]{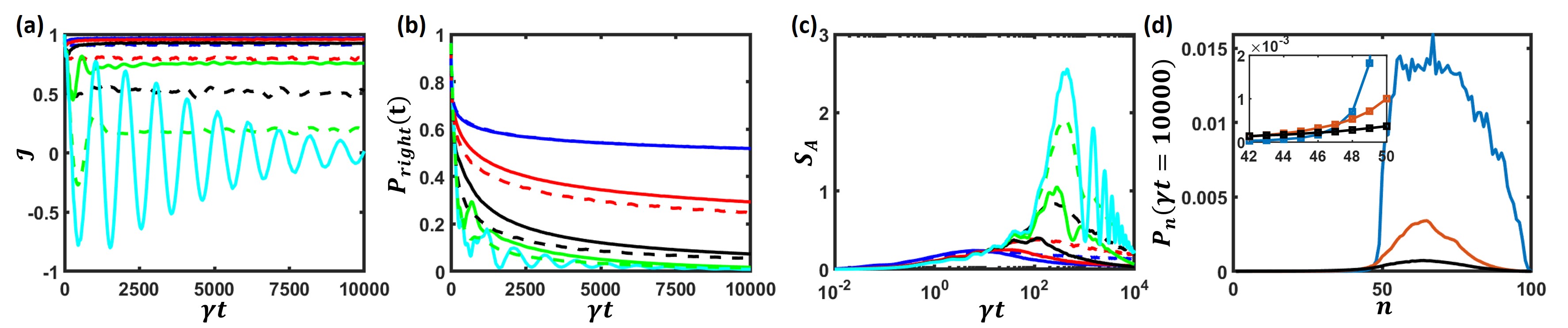}
\par\end{centering}
\caption{\label{fig:system} Schematic plot for an emitter-waveguide interface
with chiral couplings ($\gamma_{L}$ and $\gamma_{R}$), where the
right half-chain is under disorders. The time evolution of (a) imbalance,
(b) the excitation population in the right half-chain, and (c) half-chain
entropy in logarithmic time scale for $N=100$, $\xi/\pi=0.25,$ $D=0.2,$
and $\bar{w}/\pi=0.5,0.2,0.1,0.05,0$ (blue, red, black, green, and
cyan line) in the clean-disordered (fully disordered) system as solid
(dash) line. (d) The excitation distribution at $\xi/\pi=0.25,D=0.2,\gamma t=10^{4}$
for $\bar{w}/\pi=0.5,0.1,0.05$ (blue, red, and black). The inset
figure is the excitation distribution near the interface.}
\end{figure*}

We consider a general model in Lindblad forms for a periodic 1D quantum
emitters coupled to the photonic waveguides with chiral couplings
\citep{Pichler_2015,Jen_2020}, as depicted in Fig.~\ref{fig:system}.
For each two-level emitter, we denote $|e\rangle$ and $|g\rangle$
for the excited and ground states with transition frequency $\omega_{eg}$.
Without any field driving, the density matrix $\rho$ of $N$ two-level
emitters evolution in the interaction picture is determined by the
master equation (See Appendix. \ref{sec:Resonant-dipole-dipole-interacti})
$(\hbar=1)$ 

\begin{equation}
\frac{d\rho}{dt}=-i[H_{L}+H_{R},\rho]+\mathcal{L}_{L}[\rho]+\mathcal{L}_{R}[\rho],\label{eq:1}
\end{equation}
where the free evolution of the energy difference $\sum_{\mu}^{N}\omega_{eg}|e\rangle_{\mu}\langle e|$
is absorbed into the excited states as we assume the resonance between
two level energy difference and the field that excites the emitters.
The coherent parts and dissipative parts in Equation. (\ref{eq:1})
read

\begin{equation}
H_{L(R)}=-i\frac{\gamma_{L(R)}}{2}\sum_{\mu<(>)\nu}\sum_{\nu=1}^{N}(e^{ik_{s}|x_{\mu}-x_{\nu}|}\sigma_{\mu}^{\dagger}\sigma_{\nu}-\textrm{H.c.}),\label{eq:2}
\end{equation}
and

\begin{align}
\mathcal{L}_{L(R)}[\rho] & =-\frac{\gamma_{L(R)}}{2}\sum_{\mu,\nu}^{N}e^{\mp ik_{s}(x_{\mu}-x_{\nu})}(\sigma_{\mu}^{\dagger}\sigma_{\nu}\rho\nonumber \\
 & +\rho\sigma_{\mu}^{\dagger}\sigma_{\nu}-2\sigma_{\nu}\rho\sigma_{\mu}^{\dagger})\label{eq:3}
\end{align}
respectively, where $\sigma_{\mu}^{\dagger}\equiv|e\rangle_{\mu}\langle g|$
with $\sigma_{\mu}=(\sigma_{\mu}^{\dagger})^{\dagger}$ denotes the
dipole operator. $k_{s}$ and $\gamma_{L(R)}$ denote the photon wave
vector and the coupling strength to the left(right) of each quantum
emitter.Eq. (\ref{eq:1}) is obtained with Born-Markov approximation
\citep{Lehmberg_1970} under one-dimensional reservoirs, where non-reciprocal
and infinite-range photon-mediated dipole-dipole interactions determine
the spin-exchange processes \citep{Dicke_1954}. 

The asymmetric of the coupling strength $\gamma_{L(R)}$ are dependent
on the photon propagation direction and the polarization of the transition
dipole moment of the quantum emitters \citep{Lodahl_2017,Bliokh_2015_2}.
When the light is strongly transversely confined near the surface
of the waveguide, the confinement results in a relation between local
polarization and the photon propagation direction.Therefore, polarization-dependent
coupling of an emitter results in the chiral coupling. Until now,
the chiral feature has been achieved on atoms coupling to a nanofiber
\citep{Scheucher_2016,Petersen_2014}, or a photonic waveguide \citep{S_llner_2015,Mitsch_2014}.

A parameter to specify the directionality of couplings $D\equiv\frac{\gamma_{R}-\gamma_{L}}{\gamma_{R}+\gamma_{L}}$
\citep{Mitsch_2014} defines the tendency of effective light transfer.
The decay rate $\gamma=\gamma_{R}+\gamma_{L}\equiv2|dq(\omega)/d\omega|_{\omega=\omega_{eg}}g_{k_{s}}^{2}L$
\citep{Gonz_lez_Tudela_2013}, where $|dq(\omega)/d\omega|$ denotes
the inverse of group velocity with a resonant wave vector $q(\omega)$,
$g_{k_{s}}$ denotes the coupling strength, and $L$ denotes the quantization
length. $D\in[-1,1]$ specifies the trend of photon exchange between
quantum emitters, and when $D=0$, the system returns to a reciprocal
coupling regime. For one-dimensional quantum emitters array with equal
interatomic distances, $\xi\equiv k_{s}|x_{\mu+1}-x_{\mu}|$ is used
to specify the photon-mediated dipole-dipole interactions strength
which mediates the whole system. An additional significant parameter
$W_{\mu}\in\pi[-\bar{w},\bar{w}]$ is introduced to denote the onsite
phase disorders with $\bar{w}\in[0,1]$. Onsite phase disorders $W_{\mu}$
can be established in a quantum emitter array from atomic deviated
positions, leading to a deviation of $\xi$ in Eqs. (\ref{eq:2})
and (\ref{eq:3}).

As shown in Fig.~\ref{fig:system}, the system is composed of the
clean zone and disordered zone. For emitters in the clean zone, there
are no onsite atomic deviated positions. Whereas additional position
fluctuations are added for emitters in the disordered zone, which
causes onsite phase disorders. The system is initialized from a single
excitation in the disordered zone. As a consequence, the single excitation
subspace composed of $|\psi\rangle_{\mu}=|e\rangle_{\mu}|g\rangle^{\otimes(N-1)}$
should be sufficient for describing the system evolution. In this
subspace, the excitation state can be written as $|\Psi(t)\rangle=\Sigma_{\mu=1}^{N}a_{\mu}(t)|\psi(t)\rangle$,
and the non-Hermitian effective Hamiltonian reads \citep{Stannigel_2012,Pichler_2015},

\begin{align}
H_{eff} & =-i\frac{\gamma_{L}}{2}\sum_{\mu<\nu}\sum_{\nu=1}^{N}e^{ik_{s}|x_{\mu}-x_{\nu}|}\sigma_{\mu}^{\dagger}\sigma_{\nu}\nonumber \\
 & -i\frac{\gamma_{R}}{2}\sum_{\mu>\nu}\sum_{\nu=1}^{N}e^{ik_{s}|x_{\mu}-x_{\nu}|}\sigma_{\mu}^{\dagger}\sigma_{\nu}-i\frac{\gamma}{2}\sum_{\nu=1}^{N}\sigma_{\nu}^{\dagger}\sigma_{\nu}.
\end{align}
Therefore, the state evolution can be reduced to 

\begin{align}
\dot{a}_{\mu}(t) & =\gamma_{L}\sum_{\mu<\nu}e^{-i(\mu-\nu)\xi-i(W_{\mu}-W_{\nu})}a_{\nu}(t)-\frac{\gamma}{2}a_{\mu}(t)\nonumber \\
 & -\gamma_{R}\sum_{\mu>\nu}e^{i(\mu-\nu)\xi-i(W_{\nu}-W_{\mu})}a_{\nu}(t).\label{eq:4}
\end{align}
In the below, we solve the time dynamics of excitation via Eq. (\ref{eq:4})
and investigate the influence of the interface on the localization
in the disordered zone.localization in disordered zone.

\section{localization in disordered zone and near interface\label{sec:local}}

First, we would like to clarify the influence of the clean zone in
the TLQE system. Consequently, we focus on the localization robustness
in the disordered zone and study the excitation dynamics. To make
sure the excitation is not influenced by the distance to the interface
and the boundary, we consider the symmetric Dicke state in the disordered
zone as the initial state. Therefore, we have $a_{\mu}(0)=0$ and
vanishing phase disorders $W_{\mu}$ in the clean zone. As a comparison,
we use a fully disordered TLQE system initialized with the symmetric
Dicke state that $a_{\mu}=1/\sqrt{N/2}$ for $\mu\in[\frac{N}{2}+1,N]$
and investigate the effect of the clean zone on the excitation delocalization.

As time evolves, the dipole-dipole interaction between emitters transports
and dissipates the excitations. To quantify the excitation transport
from the right half-chain, we define a time-dependent imbalance

\begin{equation}
\mathfrak{I}(t)\equiv\frac{\sum_{\mu>N/2}\langle|a_{\mu}(t)|^{2}\rangle-\sum_{\mu\leq N/2}\langle|a_{\mu}(t)|^{2}\rangle}{\sum_{\mu=1}^{N}\langle|a_{\mu}(t)|^{2}\rangle},
\end{equation}
where $\langle\cdot\rangle$ denotes the average values of multi-times
simulations. 

In Fig.~\ref{fig:system}(a), in the disorder-free case, the oscillations
of the imbalance reflect the multiple excitation exchanges between
two halves of the chain, and the convergence toward $\mathfrak{I}=0$
demonstrates the delocalization in the absence of disorder. When the
disordered strength is applied, in a fully disordered system, the
imbalance decreases at a short time, indicating the excitation transport
to the left half-chain. The subsequent revival is a result of excitation
exchange and dissipation in the left half-chain. By contrast, in a
clean-disordered system, the imbalance converges to a larger value,
indicating a less excitation confinement in the clean zone compared
to the fully disordered system. As $\bar{w}$ increases, the $\mathfrak{I}$
for the clean-disordered case approaches the fully disordered ones,
since excitation exchanges are suppressed in both cases.

To further investigate the localization behavior in the right half-chain,
in Fig.~\ref{fig:system}(b) we present the population remaining
in the right half-chain as time evolves. In the fully disordered TLQE
system, the remaining excitation demonstrates the sustained behaviors
in long-time dynamics. In a clean-disordered system, sustained excitation
also remains, which demonstrates that delocalization or excitation
transport does not dominate the process through spin-exchange interaction
with emitters in the clean zone. It indicates the robustness of localization
in the disordered zone, even with the presence of emitters in the
clean zone which may allow additional excitation loss processes. It
appears that the remaining excitation in the clean-disordered system
is more than the fully disordered system. However, this property depends
on the initial states and interatomic distance $\xi$. Yet, significant
excitation in the clean-disordered system still sustains under different
initial states and $\xi.$

Besides observing the excitation dynamics, the quantum correlation
between the clean zone and the disordered zone can also be quantified
through the half-chain entropy. With entanglement entropy, the information
flow and dissipation in the left half-chain can be identified. The
left half-chain entropy is defined as $S_{L}(t)\equiv\langle\mathrm{Tr}\rho_{L}(t)\mathrm{ln}\rho_{L}(t)\rangle$,
where $\rho_{L}(t)\equiv\mathrm{Tr_{\mathit{R}}}[\rho(t)]$, and $L(R)$
denotes the left (right) part of the array \citep{Bardarson_2012}.
Given that coherent spin-exchange interaction and collective decay
strength in the same order as $\gamma$, the time-evolving entropy
is mediated by these two factors across all time scales. In Fig.~\ref{fig:system}(c),
the entanglement entropy initially rises with a similar quantity during
short-time dynamics for all ranges of disorder strength since short-time
dynamics does not engage much of the disorders but deviates after
$t\sim\gamma^{-1}.$ In the strong disorder regime, the entropy reaches
a smaller peak value faster and decreases due to excitation dissipation,
indicating the difficulty of information access in the left half-chain
for both clean-disordered and fully disordered systems. In comparison,
in the disorder-free case, a larger peak value with damping oscillation
behavior of the half-chain entropy shows the excitation transport
between two half-chains, which indicates the delocalized excitation
dynamics. However, in a weaker disorder regime, the competition between
interaction and dissipation leads to different behaviors in these
setups. For the clean-disordered system, a smaller and earlier entropy
peak demonstrates that the decay process dominates in the clean zone
when $\gamma t\sim10^{2}$. In contrast, a larger and delayed entropy
peak represents a more lasting dominance of spin-exchange interaction
within this time scale, which leads to the faster loss of excitation
transport into the clean zone.

To directly examine the remaining excitation under the influence of
the clean zone after long-time evolution, we present the distribution
of sustained excitation in the clean-disordered system. In Fig.~\ref{fig:system}(d),
sustained excitation appears in the clean zone near the interface.
The non-zero imbalance shown in Fig.~\ref{fig:system} (a) also indicates
excitation localized in the clean zone. As shown in the inset plot
of Fig.~\ref{fig:system}(d), a larger disorder results in more excitations
localized near the interface, while a weaker disorder shows a larger
spread for excitation localization, which can be attributed to a balance
between the disorder-assisted localization from the disordered zone
and the clean-zone-assisted delocalization. The feature that finite
excitation localizes at the interface in the clean zone suggests the
strong effect as well as the robustness of localization in the disordered
zone. From the results in Fig.~\ref{fig:system}, we find that even
though dipole-dipole interaction allows excitation hopping from the
disordered zone to the clean zone, the excitation in the disordered
zone is still localized as time evolves, which implies the robustness
of localization of TLQE system in this parameter regime.

\section{Level statistics and Participation Ratio\label{sec:Participation-Ratio}}

\begin{figure}[t]
\begin{centering}
\includegraphics[scale=0.3]{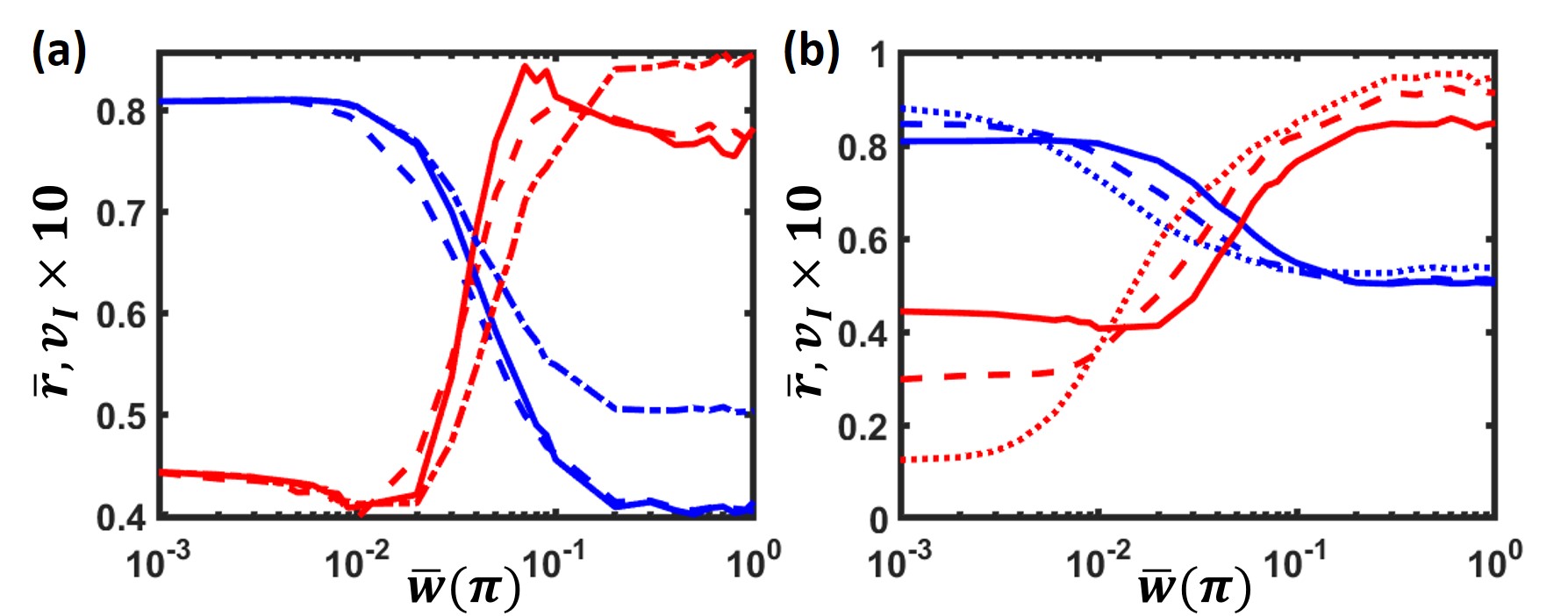}
\par\end{centering}
\caption{\label{fig:gap} The mean gap ratio and the intrasample variance with
$N=100$ and $D=0$ for different disorder strength $\bar{w}$. (a)
At $\xi/\pi=0.5$, we plot the mean gap ratio (blue line) and intrasample
variance (red line) of the fully disordered system (dash line), clean-disordered
system (dash-dot line), and the clean-disordered system composed of
eigenvalues with more excitation in disordered zone (solid line).
(b) The mean gap ratio (blue line) and intrasample variance (red line)
of clean-disordered system at $\xi/\pi=0.125$ (dot), $0.25$ (dash),
$0.5$ (solid).}
\end{figure}

To identify the transition from excitation localization to delocalization,
we investigate additional measures that help determine the localization
crossover, which are level statistics and participation ratio. Level
statistics has been extensively applied in studies of quantum chaos
\citep{Abanin_2019}, which is highly related to the thermalization
properties of the system and allows for the classification of them
based on the distribution of energy levels. In the tight-binding model
of interacting fermions \citep{Oganesyan_2007,Sierant_2019}, the
gap ratio presents Gaussian orthogonal ensemble (GOE) or Poisson statistics
(PS) in clean or disordered conditions, respectively. The gap ratio
$r_{j}$ is defined by the ascendant eigenenergy $E_{j}$ and the
adjacent gaps $\delta_{j}\equiv E_{j+1}-E_{j}$, where $r_{j}\equiv$
min$\{\delta_{j},\delta_{j-1}\}/$max$\{\delta_{j},\delta_{j-1}\}$
\citep{Oganesyan_2007}. For each disorder realization, $r_{a}\equiv\sum_{j=2}^{N-1}r_{j}/(N-2)$
is defined and the mean gap ratio is obtained by $\bar{r}=\langle r_{a}\rangle.$
A level repulsion in $\bar{r}_{GOE}\thickapprox0.53$ is presented,
compared to $\bar{r}_{PS}\thickapprox0.39$ with uncorrelated energy
levels due to strong disorder. The intrasample variance $v_{I}\equiv\Sigma_{n=2}^{N-2}\langle r_{n}^{2}-r_{a}^{2}\rangle$
can also be calculated, which shows the fluctuations of level repulsions
\citep{Sierant_2019}. A larger level repulsion indicates the correlation
between energy level spacings, which is a crucial feature of the clean
system. On the contrary, the disordered system presents a larger intrasample
variance with a smaller gap ratio. Through the eigenspectrum of photon-mediated
dipole-dipole interaction between each emitter, the level statistics
obtained in this TLQE system can provide distinctive features of the
localization.

Therefore, via the coupling matrix in Eq. (\ref{eq:4}), we can extract
information from the energy levels of the clean-disordered system.
We note that the eigenvalues $\lambda_{n}$ are complex due to the
non-Hermiticity of the TLQE system, where the real and imaginary parts
of eigenvalues correspond to the collective decay rates and energy
shifts, respectively. To make sure that the level statistics can be
used as a measure for localization crossover, we select the valid
sectors of energy levels when the ratio between resonance widths and
energy spacings, $(-\mathrm{Re}[\lambda_{n}+\lambda_{n+1}]/2)/(\mathrm{Im}[\lambda_{n+1}-\lambda_{n}])$,
is less than $1/2$ and obtain the gap ratios and intrasample variances.
In this way, we can exclude the sectors with superradiant decay rates
and retain the eigenspectrum which mostly contains the subradiant
sector of decay rates. 

In Fig.~\ref{fig:gap}(a), we present the level repulsion and fluctuation
at $D=0$. As the disorder increases, the mean gap ratio decreases
and crosses over from delocalization to localization. Based on this,
we define the mid-point of the mean gap ratio as the crossover points
of disorder strength toward localization. In strong disorder regime,
the clean-disordered system exhibits a stronger level repulsion compared
to the fully disordered system due to the contribution of the clean
zone. This is because the eigenvectors mostly composed of single excitation
in the clean zone still contain the feature of an clean system, and
therefore these corresponding eigenvalues contribute to a larger mean
gap ratio. To examine the feature of $\bar{r}$ in the disordered
system, we choose the eigenvector components containing single excitation
in the disordered zone more than $0.25$ as an additional constraint
for selecting eigenvalue sectors. With this constraint, the mean gap
ratio and intrasample variance return to the values similar to those
of the fully disordered system, indicating the influence of the clean
zone on the level statistics, which coincides with the excitation
localization in disordered zone shown in Fig.~\ref{fig:system}(d). 

\begin{figure}[t]
\begin{centering}
\includegraphics[scale=0.275]{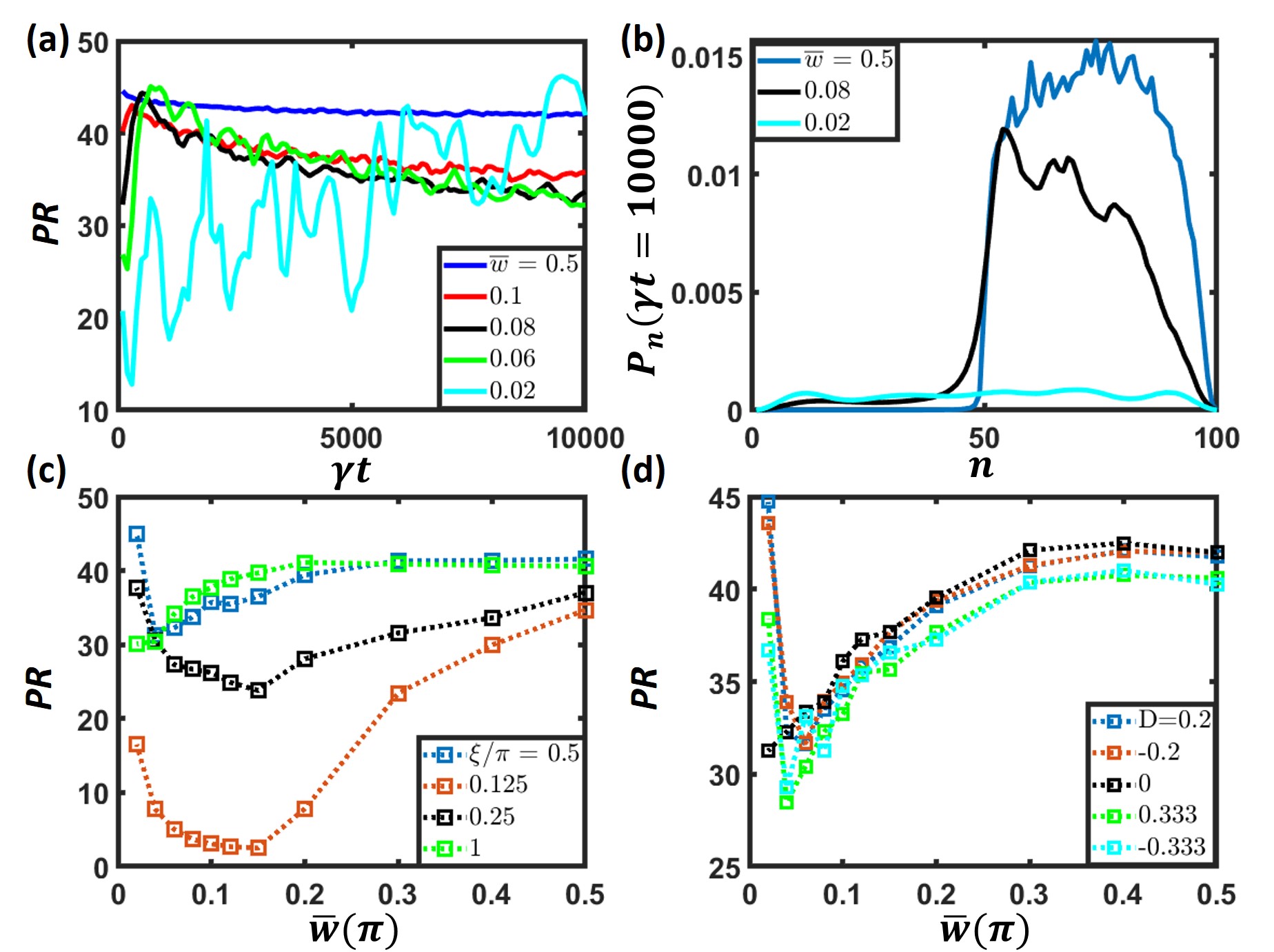}
\par\end{centering}
\caption{\label{fig:pr} Participation ratio for $N=100$ with equal sizes
of clean zone and disordered zone. (a) Time evolution of PR for $\xi=0.5\pi,$
$D=0.2$ under different disorder strength $\bar{w}$. (b) The excitation
distribution for localization with $\bar{w}=0.5$, $0.08$, and $0.02$.
The PR (c) for $D=0.2$ at $\xi/\pi=0.125,0.25,0.5,1$ and (d) with
$\xi=0.5\pi$ at $D=0,\pm0.2,\pm0.333$ at $\gamma t=10000$.}
\end{figure}

To further study the effect of dipole-dipole interactions on the level
statistics, in Fig.~\ref{fig:gap}(b), we show the mean gap ratios
and the intrasample variance for different $\xi$. As $\xi$ increases,
the crossover points of disorder strength toward localization shift
to a stronger value, which suggests a facilitation of the excitation
delocalization, similar to the fully disordered case \citep{Jen_2021}.
This can be attributed to the dominance of collective decays at a
small $\xi$ and sustained correlations among emitters at a large
$\xi$. The former leads to decoherences of the system and favors
the effect of disorders, while the latter indicates a competition
between dipole-dipole interactions and disorders, pushing the crossover
points toward a stronger disorder strength. We note that at $\xi=0$
and $\pi$ in the reciprocal coupling case, decoherence-free states
emerge and the localization of atomic excitations sustain indefinitely
without disorders.

However, the level statistics can only be utilized in the reciprocal
regime. To examine the localization crossover in the non-reciprocal
regime $(D\neq0)$, we introduce participation ratio (PR) \citep{Murphy_2011}
as a measure to determine the excitation distribution in comparison
to the uniform distribution. Since the excitation loss is inherent
to the TLQE system, the probability amplitude of the system decreases
over time. To focus on the distribution of the remaining excitation,
time-dependent PR is defined with $\tilde{P}_{\mu}(t)$, which represents
the onsite probability with the sum normalized to $1$ 

\begin{equation}
\mathrm{PR}(t)\equiv\frac{(\sum_{\mu=1}^{N}\langle\Delta\tilde{P}_{\mu}(t)\rangle)^{2}}{\sum_{\mu=1}^{N}(\langle\Delta\tilde{P}_{\mu}(t)\rangle)^{2}},
\end{equation}
where $\Delta\tilde{P}_{\mu}(t)\equiv|\tilde{P}_{\mu}(t)-N^{-1}|\Theta(\tilde{P}_{\mu}(t)-N^{-1})$
with Heaviside function $\Theta$. 

As shown in Fig.~\ref{fig:pr}(a), PR is close to the number of quantum
emitters in the disordered zone $N/2$ under strong disorders. Owing
to the strong localization from the disorders, the probability amplitude
is confined in the disordered zone except at the end. As the disorder
strength decreases, PR gradually decreases, which also reflects in
the wider excitation spread at the interface shown in Fig.~\ref{fig:pr}(b).
We note that PR can serve as a measure for excitation localization
when the system is deep in the localization side, but it cannot be
used to determine the crossover behavior to localization since it
fluctuates over time in the delocalized side. 

In Fig.~\ref{fig:pr}(c), we further explore how PR changes under
different dipole-dipole interaction strength $\xi$ and directionality
$D$. In strong disorder regime, at $\xi/\pi=0.125$, a significant
decreasing trend of PR occurs as $\overline{w}$ decreases, which
demonstrates the tendency of excitation delocalization. On the contrary,
at $\xi/\pi=1$, a less significant decrease of PR occurs as $\overline{w}$
decreases, which represents that the localization is still sustained
at a smaller disorder. We observe a similar behavior in PR at $D=0$,
that a smaller $\xi$ facilitating delocalization, which is contrary
to the analysis from level statistics in Fig.~\ref{fig:gap}(b).
This deviation can be attributed to the initial-state dependence of
PR, whereas level statistics in general does not depend on it. In
Fig.~\ref{fig:pr}(d), the influences of directionality $D$ are
demonstrated. When $D=0$, PR is larger in the localization side due
to the absence of a propagation tendency for the excitation. As $D$
increases, smaller PR reflects a weaker localization since a larger
directional coupling strength evades the influence of the disorders
\citep{Jen_2020}. Meanwhile, the trending of PR is similar for the
same $D$ with a different sign, demonstrating that the tendency of
excitation exchange direction does not modify the localization behavior
near the clean-disordered interface even under asymmetric initial
states. Even though PR is irregular in the delocalization side, its
variation still reflects some features of delocalization in the localization
regime in a broad parameter range.

\section{Directional Photon loss ratio\label{sec:Directional-Photon-loss}}

Besides investigating the localization from the remaining excitation
through PR, we study the photon loss from the ends of the TLQE, which
can be experimentally measurable by detecting photons dissipation
from the ends. Via the input-output theory \citep{Collett_1984,Gardiner_1985,Lalumi_re_2013},
we can illustrate the relationship between the output signal from
both ends of the systems and the atomic response. Starting with Hamiltonian
including photon basis (Appendix~\ref{sec:Photon-Flux}), the photonic
output field is derived as

\begin{equation}
\langle a_{out,d}^{\dagger}(t)a_{out,d}(t)\rangle=\langle\sum_{\mu=1}^{N}\gamma_{d}e^{m_{d}ik(r_{\mu}-r_{\nu})}\sigma_{\mu}^{\dagger}(t)\sigma_{\nu}(t)\rangle,
\end{equation}
where $a_{out,d}^{\dagger}$ denotes the creation operator of photons
leaving the system, with photon propagation direction $d$ such that
$m_{d}=+1(-1)$ for photon toward the left (right). $\gamma_{d}=\gamma_{L(R)}$
quantifies the strength for left (right) propagating photon coupling.
Besides the collective interaction between emitters, the position
of the excitation also determines the loss from the system. If the
excitation is close to the left (right) end, then the excitation is
more likely to escape from the system as a photon propagating toward
the left (right). Therefore, under the asymmetric initial system setup,
competing between photon loss from the end and delocalization from
the clean-disordered interface can be measured by the directional
photon loss ratio (DPLR)

\begin{equation}
\mathrm{DPLR}(t)\equiv\frac{\int_{0}^{t}\langle a_{out,R}^{\dagger}(\tau)a_{out,R}(\tau)\rangle d\tau}{\int_{0}^{t}\langle a_{out,L}^{\dagger}(\tau)a_{out,L}(\tau)\rangle d\tau}.
\end{equation}

\begin{figure}
\begin{centering}
\includegraphics[scale=0.275]{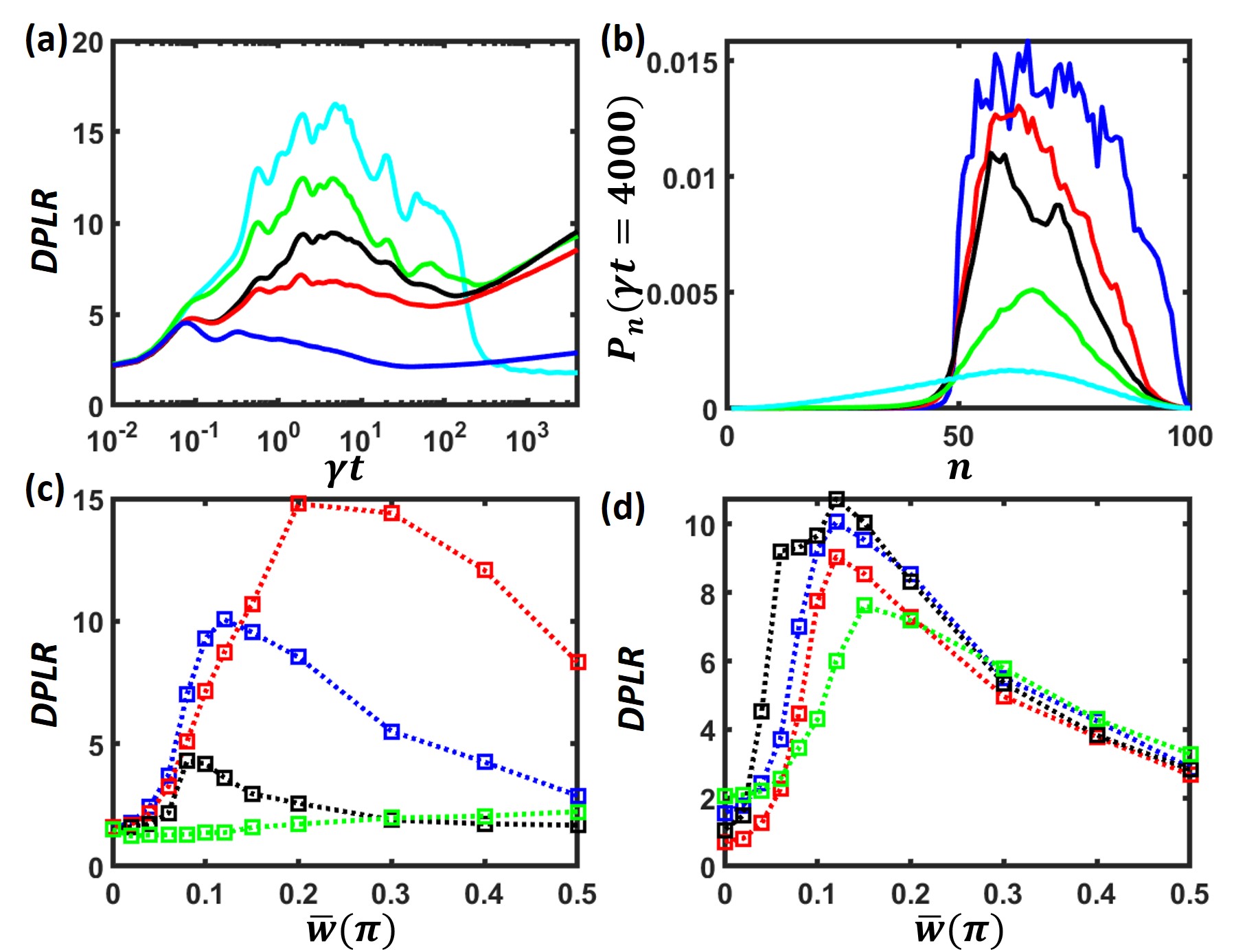}
\par\end{centering}
\caption{\label{fig:DPLR}Directional photon loss ratio for $N=100$ with equal
sizes of clean zone and disordered zone. (a) Time-evolved DPLR and
(b) the probability distribution when $\bar{w}=0.5$ (blue line),
$0.2$ (red line), $0.15$ (black line), $0.1$ (green line), $0.02$
(cyan line) with $\xi=0.25\pi,D=0.2$. (c) DPLR with $D=0.2$ for
$\xi/\pi=0.125$ (red line), $0.25$ (blue line), $0.5$ (black line),
$1$ (green line) at $\gamma t=4000$. (d) DPLR with $\xi=0.25\pi$
for $D=-0.2$ (red line), $0$ (black line), $0.2$ (blue line), $0.333$
(green line) at $\gamma t=4000$.}
\end{figure}

In Fig.~\ref{fig:DPLR}(a), we show DPLR under different disorder
strength $\bar{w}$ with $\xi=0.25\pi$ as an example. In short-time
dynamics, the DPLR rises with a similar quantity for all ranges of
disorder strength since the disorders are not engaged in short-time
dynamics. However, the evolution of DPLR deviates after $\gamma t\sim10^{-1}$.
In strong disorder regime, the DPLR only slightly decreases until
$\gamma t\sim10^{1}$ and increases after that. The decrease of DPLR
is due to the excitation transport into the clean zone. Once the excitation
transport to the clean zone, most of the excitation will exit from
the left end of the system, lowering DPLR. After most of the excitation
transport into the clean zone is lost, the loss from the sustained
excitation in the disordered zone dominates the decay process, resulting
in the subsequent revival of DPLR. As a comparison, in a weaker disorder
regime, a larger and delayed peak and subsequent decline of DPLR represent
the tendency of delocalization. The former is owing to larger direct
loss from the right end, which is also demonstrated in Fig.~\ref{fig:DPLR}(b),
and the latter results from more excitation transport into the clean
zone. After $\gamma t\sim10^{2}$, in weak disorder regime ($\bar{w}\sim[0.1,0.2]$),
when the localization is still sustained, the excitation transport
into the clean zone is limited, resulting in larger DPLR after long-time
evolution. On the contrary, when the excitation is delocalized $(\bar{w}\lesssim0.1)$,
the excitation can transport into the left half-chain and exits from
the left end. Therefore, DPLR decreases and approaches to $\gamma_{R}/\gamma_{L}$
as the disorder strength decreases to zero in long-time dynamics.
Accordingly, we define the localization crossover point of $\bar{w}$
at which the maximal DPLR appears in the long-time regime.

\begin{figure*}[!t]
\centering{}\includegraphics[scale=0.3]{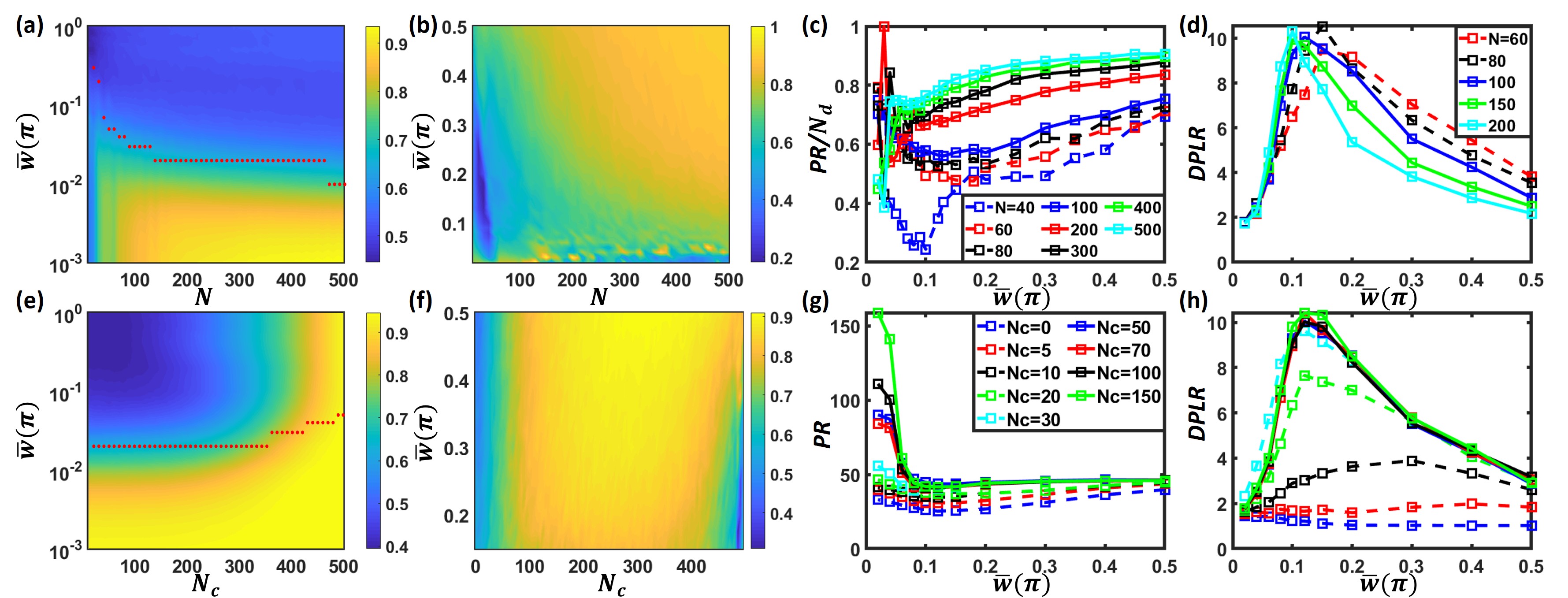}\caption{\label{fig:size}The mean gap ratio, PR and DPLR under different sizes
and clean-disordered zone ratios. (a) The mean gap ratio under $D=0$,
(b) $PR/N$ under $D=0.2,\gamma t=4000$ with different quantum emitter
numbers and $\bar{w}$ for $\xi=0.25\pi$ and $N_{c}=N_{d}.$ (c)
$PR/N_{d}$ with different $N$ in (b). (d) DPLR for $N=50,100,150,200$
under different disorder strength with equal size in clean order zone
and disordered zone size. (e) The mean gap ratio with $D=0$ (f) PR
with $D=0.2,\gamma t=4000$ varies $N_{c}$ and $\bar{w}$ while $N=500$
and $\xi=0.25\pi$. (g)(h) PR and DPLR with different size of clean
zone while $N_{d}=50$, $\xi=0.25\pi,D=0.2,\gamma t=4000.$ The red
dots in (a) and (e) represents the localization crossover points.}
\end{figure*}

To further investigate the influence of the dipole-dipole interaction
strength and directionality $D$, in Fig.~\ref{fig:DPLR}(c), we
present DPLR under different $\xi$. The rising and decreasing behavior
of DPLR is common when $\xi/\pi=0.125,0.25,0.5$. Larger localization
crossover points appear at smaller $\xi$ also indicates the facilitation
of delocalization at smaller $\xi$. However, when $\xi=\pi$, there
is no obvious rising or decreasing in DPLR under all ranges of disorder
strength. In this case, the initial symmetric Dicke state is close
to the decoherence-free state at $D=0$. Therefore, the excitation
dynamics at $\xi=\pi$ is different from the other $\xi$, and crossover
point is hard to be determined in this condition. For the effect of
$D$, in Fig.~\ref{fig:DPLR}(d), DPLR under different directionality
$D$ is presented. In strong disorder regime, the value of DPLR is
similar since the excitation dynamics is mainly determined by the
disorders. Yet, as $\bar{w}$ decreases, the behavior deviates for
different $D.$ The crossover point occurs at a larger $\bar{w}$
when $|D|$ is larger, showing that the excitation localization emerges
with a larger disorder strength to counteract the effect of $D$.
In comparison, near the disorder-free regime, the excitation dynamics
is mainly determined by directionality $D$, and DPLR converges to
$\gamma_{R}/\gamma_{L}$ for each $D.$ DPLR serves as an experimental
measure for delocalization by utilizing the inherent loss of TLQE
systems, which also reflects the complex interplay of disorders, dipole-dipole
interaction, and directionality. We note that the photon loss also
depends on the initial state. Therefore, different initial states
result in different behaviors of the DPLR.

\section{Size effects\label{sec:Size-effects}}

Finally, we search for the signature of quantum avalanche-like behaviors
in the TLQE system through investigating the size effects on the clean-disordered
interface via the mean gap ratio, PR, and DPLR. The size of the clean
zone plays a crucial role in determining quantum avalanche dynamics
in many-body localization system \citep{L_onard_2023}, where a larger
clean zone size leads to a faster delocalization from the clean-disordered
interface. Therefore, we explore the delocalization phenomenon in
a similar fashion by modifying the relative sizes between the clean
and the disordered zones. In Fig.~\ref{fig:size}, we present several
cases of different size parameters. The first case is to change the
overall size of the TLQE systems with a fixed ratio of clean zone.
The results are shown in Fig.~\ref{fig:size}(a). As the overall
system size increases, the phase crossover points measured by the
mean gap ratio occur at smaller disorder strength $\bar{w}$, which
demonstrates the enhancement of localization via scaling. In addition,
Figs.~\ref{fig:size}(b) and \ref{fig:size}(c) present that the
value of $PR/N_{d}$ decreases slower as $\overline{w}$ decreases
in the strong disorder regime, where $N_{d}(N_{c})$ denotes the atomic
number in the disordered (clean) zone. It indicates that the robustness
of the localization is also enhanced in a larger system. Meanwhile,
DPLR in Fig.~\ref{fig:size}(d) shows that the localization crossover
points and the photon scattering behavior is saturated as size increases,
which indicates that the system reaches a thermodynamic limit of scaling.
Both $PR/N_{d}$ and DPLR provide insights into the finite size effects
on the clean-disordered interface and the influence of direct loss
at the end of the system. 

The second case is to investigate the effect of tuning the ratio of
the clean-disordered zone on the phase crossover of excitation localization.
As shown in Fig.~\ref{fig:size}(e), as the ratio of the clean zone
increases, the phase crossover points determined by the mean gap ratio
occur at larger disorder strength $\bar{w}$, which demonstrates that
the ratio of the clean zone facilitate the excitation delocalization.
In Fig.~\ref{fig:size}(f), PR over the atomic number in disordered
zone $N_{d}$ is presented for different clean-disordered zone size
ratios when $N=500$. We find that as the ratio of clean zone increases,
the value of $PR/N_{d}$ decreases faster as $\overline{w}$ decreases
in the strong disorder regime. We note that when $N_{d}$ is close
to $N$, the excitation distribution will be close to the uniform
distribution. Therefore, PR is inaccurate in this parameter regime
since PR involves a comparison of excitation distribution to the uniform
distribution. The results shown in Figs.~\ref{fig:size}(e) and \ref{fig:size}(f)
represent the enhancement of excitation delocalization as the ratio
of clean zone increases. In other words, as the ratio of disordered
zone increases, the excitation localization is favored. On the other
hand, as the overall size of the system increases with a fixed ratio,
the excitation localization is favored for a smaller disorder strength
owing to the multi-atom effect on the suppression of spin hoppings.

In the end, we follow the analysis in the experimental work \citep{L_onard_2023}
by tuning the number of emitters in the clean zone $N_{c}$ while
keeping the number of emitters in the disordered zone $N_{d}$ constant.
As shown in Fig.~\ref{fig:size}(g), in strong disorder regime, when
$N_{c}<20$, PR decreases more significantly as the disorder strength
$\overline{w}$ decreases due to the boundary loss from the left end.
In contrast, when the clean zone size is large enough, PR converges
to a constant value in the strong disorder regime, which reflects
the excitation localization in the disordered zone at the interface.
Meanwhile, in Fig.~\ref{fig:size}(h), when $N_{c}<20$, DPLR remains
almost constant under every disorder strength because of excitation
loss from both ends of the TLQE system. Whereas once $N_{c}\apprge20$,
the phase crossover points determined by DPLR becomes constant, and
the rising and declining features of DPLR are recovered, which demonstrates
the convergence of collective dissipation behaviors once the boundary
loss from the left end is not dominated. These suggest that the effect
of overall size on enhancing the localization is more significant
than the effect of the clean zone ratio that facilitates delocalization
under a fixed number of the whole chain. 

\section{Discussion\label{sec:Conclusion}}

The absence of delocalization enhancement in the TLQE system demonstrates
a fundamental difference in localization mechanisms compared to other
systems. In TLQE system, the position fluctuations of the emitters
result in a random and complex spin-exchange process between each
emitter, which leads to excitation localization. Under this disorder-assisted
localization, the excitations still redistribute through the coherent
hoppings and collective decay as time evolves. From the analysis of
level statistics, we find that while TLQE exhibits level repulsion
under weak disorder, larger mean gap ratio appears in clean-disordered
system under strong disorder regime compared to fully disordered system
due to the contribution of the clean zone. We further calculate the
participation ratio which can quantify the distribution of excitations
and localization behaviors. In addition, the directional photon loss
ratio also reflects the localization behaviors. When the excitations
are delocalized, the excitations spread through the whole chain and
decay from both sides, resulting in the declining feature of DPLR
in the delocalized side. Therefore, we utilize the maximal point of
DPLR as a measure of localization crossover points. With these measures,
how the localization behaviors influenced by the scaling size can
be determined. A larger disordered zone leads to the tendency of localization,
whereas increasing the size of the clean zone does not enhance delocalization.
This can also be interpreted as that the scaling overall size that
enhances localization surpasses the effect from the clean zone ratio
that facilitates the delocalization with a fixed overall size. 

There are several potential platforms to experimentally observe the
excitation dynamics in the clean-disordered system and detect the
directional photon loss. One potential platform is atoms coupled to
an optical fiber \citep{Corzo_2019,Prasad_2020,Solano_2017_2,Solano_2017,Sayrin_2015}.
In recent experiments \citep{Corzo_2019,Prasad_2020,Solano_2017_2,Solano_2017,Sayrin_2015},
hundreds of cesium atoms can be confined surrounding an optical fiber
via the evanescent field of the guided mode. Dipole-dipole interaction
between atoms are mediated by the guided photons in the fibers and
give rise to strong collective interaction behaviors. If photon detectors
are attached at the ends of the optical fiber, radiative photon loss
can be further detected. An alternative approach to observe the localization
to delocalization crossover can be done by directly detecting excitation
population. Through shining the laser with proper transition frequencies
from the side, the excitation can be detected. This method is widely
used in ultracold atom experiments for quantum state detection. Another
potential platform is superconducting qubits, in which many-body localization
has already been experimentally explored \citep{Roushan_2017,Xu_2018}.
With the cavity, microwave photons emitted from superconducting qubits
can be captured and observed \citep{Lu_2021,Helmer_2009,Johnson_2010}. 

Beyond the scope of this work, we propose other potential future research
directions. For example, we can extend to multiple excitations regime
for investigating how many-body effects influence current results
in this unique system. Very recently, many-body localization in waveguide
QED has been explored \citep{Manasi_2018,See_2019}. Further exploration
with input fields on the current system can provide more insights
on the excitation dynamics. Another potential work is to decompose
an atomic chain into several chains with different interatomic distances.
With the excitation transport to the interface between these dissimilar
chains, one can study the excitation transmission and reflection resulting
from the collective interactions among the emitters.

\section*{Acknowledgments}

We acknowledge support from the National Science and Technology Council
(NSTC), Taiwan, under Grants No. 112- 2112-M-001-079-MY3, No. NSTC-112-2119-M-001-
007, No. 112-2811-M-002-067, and No. 112-2112-M-002-001. We are also
grateful for support from TG 1.2 of NCTS at NTU.

\appendix
\renewcommand{\appendixname}{APPENDIX}

\section{Resonant dipole-dipole interaction in one-dimensional reservoir\label{sec:Resonant-dipole-dipole-interacti}}

We consider an emitter arrays coupled to a waveguide which offers
a common 1D reservoir \citep{Lehmberg_1970,Jen_2020_2,Gonz_lez_Tudela_2013}.
For the $\mu$-th two-level emitter, we denote $|e\rangle_{\mu}$
and $|g\rangle_{\mu}$ for the excited and ground states with transition
frequency $\omega_{eg}$. The system Hamiltonian $H_{sys}$ reads, 

\begin{align}
H_{sys} & =\hbar\sum_{\mu=1}^{N}\omega_{eg}\sigma_{\mu}^{\dagger}\sigma_{\mu}^{-},
\end{align}
where $\sigma_{\mu}^{\dagger}\equiv|e\rangle_{\mu}\langle g|$ with
$\sigma_{\mu}=(\sigma_{\mu}^{\dagger})^{\dagger}$ denotes the dipole
operator. 

The corresponding reservoir Hamiltonian $H_{r}$ reads

\begin{equation}
H_{r}=\sum_{q}\sum_{d=L,R}\hbar\omega a_{\omega_{q},d}^{\dagger}a_{\omega_{q},d},
\end{equation}
where $a_{\omega_{q},d}$are bosonic annihilation operators for the
right (left) propagating bath modes $q$ with frequency $\omega_{q}$
for $d=R(L)$.

In the interaction picture, the system-reservoir interaction is given
by

\begin{align}
H_{sys-r} & =i\hbar\sum_{\mu,d,q}g_{\omega_{q},d}(\sigma_{\mu}^{\dagger}+\sigma_{\mu})\nonumber \\
 & (a_{\omega_{q},d}^{\dagger}e^{i(\omega_{q}t-\omega x_{\mu}/v_{d})}+a_{\omega_{q},d}e^{-i(\omega_{q}t-\omega x_{\mu}/v_{d})}),
\end{align}
where $g_{L}$($g_{R}$) is the emitter-photon coupling strength into
the left(right) propagating reservoir modes with speed $v_{d}$.

Therefore, the reservoir operators $a_{\omega_{q},d}(t)$ can be obtained
by solving the Heisenberg equations of motion 

\begin{align}
a_{\omega_{q},d}(t) & =a_{\omega_{q},d}(0)e^{-i\omega t}+i\int_{0}^{t}dt'\sum_{\mu=1}^{N}\nonumber \\
 & g_{\omega_{q},d}[\sigma_{\mu}(0)+\sigma_{\mu}^{\dagger}(0)]e^{-i\omega_{q}t'-i\omega_{q}x_{\mu}/v_{d}}.\label{eq:13}
\end{align}
For an arbitrary emitter operator $A$ acting on the Hilbert space
of the spins, we obtain

\begin{align*}
\dot{A}(t) & =i\omega_{eg}\sum_{\mu}[\sigma_{\mu}^{\dagger}\sigma_{\mu},A]-i\sum_{\mu,d,q}g_{\omega,d,q}\{e^{ik_{d,q}x_{\mu}}\\
 & [\sigma_{\mu}^{\dagger}+\sigma_{\mu},A]a_{\omega_{q},d}(t)+e^{-ik_{d,q}x_{\mu}}a_{\omega_{q},d}^{\dagger}(t)[A,\sigma_{\mu}^{\dagger}+\sigma_{\mu}]\},
\end{align*}
where $k_{d,q}\equiv\omega_{q}/v_{d}$. We assume Born-Markov approximation
is valid in our system. The Born approximation states that the coupling
between the emitters and the reservoir modes is so weak that the system
operators can be approximately related to the first order of the coupling
strength. Markov approximation suggests that the correlation time
between the bath and the system is so short compared to the system
evolution. Besides, we neglect the time retardation effect from photon
propagation, which is valid when the time scales of system operators
evolution is much larger than the time for photon propagation in the
waveguide. The initial bath state is assumed to be the vacuum state.
We then obtain the expectation value of system operator $\langle A(t)\rangle=\textrm{Tr}\{A(t)W(0)\}$
by tracing over the reservoir modes, where $W(0)$ is the density
matrix of the initial system $=\rho(0)\otimes|vac\rangle\langle vac|$.
The equation of motion $\langle A(t)\rangle$ is given by

\begin{align}
\langle\dot{A}\rangle(t) & =\sum_{\mu\neq\nu,d}i\Omega_{\mu,\nu,d}[\sigma_{\mu}^{\dagger}\sigma_{\nu},A]+\sum_{d}\mathcal{L}_{d}(A),
\end{align}
where

\begin{equation}
\mathcal{L}_{d}(A)=\sum_{\mu,\nu}\Gamma_{\mu,\nu,d}[\sigma_{\mu}^{\dagger}A\sigma_{\nu}-\frac{1}{2}(A\sigma_{\mu}^{\dagger}\sigma_{\nu}+\sigma_{\mu}^{\dagger}\sigma_{\nu}A)].
\end{equation}
The coupling strength $J_{\mu,\nu,d}\equiv\frac{1}{2}\Gamma_{\mu,\nu,d}+i\Omega_{\mu,\nu,d}$
can be obtained as

\begin{align*}
J_{\mu,\nu,d} & =\sum_{q}\int_{0}^{t}ds|g_{\omega_{q},d}|^{2}e^{ik_{d,q}x_{\mu,\nu}}(e^{i(\omega_{eg}-\omega)s}+e^{-i(\omega_{eg}+\omega)s})\\
 & =\sum_{q}|g_{\omega_{q},d}|^{2}e^{ik_{d,q}x_{\mu,\nu}}[\pi\delta(\omega_{eg}-\omega)+\pi\delta(\omega_{eg}+\omega)\\
 & +i\mathcal{P}(\omega_{eg}-\omega)^{-1}-i\mathcal{P}(\omega_{eg}+\omega)^{-1}],
\end{align*}
where $\mathcal{P}$ denotes the principal values, and $x_{\mu,\nu}\equiv x_{\mu}-x_{\nu}$.

Considering the summation of the mode to the continuous limit, $\Sigma_{q}\rightarrow\int\frac{dq}{2\pi}L$,
where $L$ is the quantization length.

Therefore, we have

\begin{align*}
J_{\mu,\nu,d} & =\int_{0}^{\infty}\frac{dq_{d}}{2\pi}\bar{g}_{\omega_{q},d}^{2}Le^{ik_{\omega,d}x_{\mu,\nu}}[\pi\delta(\omega_{eg}-\omega)\\
 & +\pi\delta(\omega_{eg}+\omega)+i\mathcal{P}(\omega_{eg}-\omega)^{-1}-i\mathcal{P}(\omega_{eg}+\omega)^{-1}]\\
 & =\int_{0}^{\infty}\frac{d\omega}{2\pi}|\partial_{\omega}q(\omega)|\bar{g}_{\omega_{q},d}^{2}Le^{ik_{\omega,d}x_{\mu,\nu}}[\pi\delta(\omega_{eg}-\omega)\\
 & +\pi\delta(\omega_{eg}+\omega)+i\mathcal{P}(\omega_{eg}-\omega)^{-1}-i\mathcal{P}(\omega_{eg}+\omega)^{-1}]\\
 & =\int_{0}^{\infty}\frac{d\omega}{4\pi}\gamma_{d}e^{ik_{\omega,d}x_{\mu,\nu}}[\pi\delta(\omega_{eg}-\omega)\\
 & +\pi\delta(\omega_{eg}+\omega)+i\mathcal{P}(\omega_{eg}-\omega)^{-1}-i\mathcal{P}(\omega_{eg}+\omega)^{-1}].
\end{align*}
where $\gamma_{d}\equiv2|\partial_{\omega}q(\omega)|\bar{g}_{\omega_{q},d}^{2}L$
and $k_{\omega,d}\equiv\omega/v_{d}$.

Finally, we simplify the equation above as 

\begin{align}
J_{\mu,\nu,d} & =\frac{\gamma_{d}}{4}e^{ik_{\omega,d}x_{\mu,\nu}}-i\frac{\mathcal{P}}{2\pi}\int d\omega\frac{\gamma_{d}e^{ik_{\omega,d}x_{\mu,\nu}}}{\omega-\omega_{eg}},\nonumber \\
 & =\frac{\gamma_{d}}{4}e^{ik_{\omega,d}x_{\mu,\nu}}+i\frac{\gamma_{d}}{4}\textrm{sin}(|k_{\omega,d}x_{\mu,\nu}|).
\end{align}
This coupling strength $J_{\mu,\nu,d}$ provides the foundation of
the chiral coupling in Eqs.(\ref{eq:1}) to Eqs.(\ref{eq:3}).

\section{Photon Flux\label{sec:Photon-Flux}}

To observe the photon flux propagating out the system, instead of
using the master equation and effective Hamiltonian \citep{Pichler_2015}
which traces out the photonic fields, we consider the Hamiltonian
including both atomic and photonic basis. Therefore, the Hamiltonian
of this system $H$ can be decomposed of the emitter term $H_{em}$,
photonic terms $H_{p}$, and emitter-photon interaction terms $H_{em-p}$:

\begin{equation}
H=H_{em}+H_{p}+H_{em-p},
\end{equation}
where 

\begin{align}
H_{em} & =\sum_{\mu}\hbar\omega_{eg}\sigma_{\mu}^{\dagger}\sigma_{\mu}^{-}\\
H_{p} & =\sum_{d}\int\hbar\omega a_{\omega,d}^{\dagger}a_{\omega,d}d\omega\\
H_{em-p} & =i\hbar\sum_{\mu,d}\int d\omega(g_{d}e^{m_{d}ik_{s}x_{\mu}})\sigma_{\mu}a_{\omega,d}^{\dagger}+\textrm{H.c.},
\end{align}
where $k_{s}$ and $g_{d}$ denote the photon wave vector and the
coupling constant between quantum emitters. The photon with propagation
direction $d$ toward the left (right) such that $m_{d}=+1(-1)$.
$a_{\omega_{d}}(a_{\omega_{d}}^{\dagger})$ denotes the annihilation
(creation) operator of the photon with frequency $\omega$ and direction
$d$ .

To obtain the output photon flux, we calculate the annihilation operator
in Heisenberg picture.

\begin{align}
\frac{da_{\omega,d}}{dt} & =\frac{i}{\hbar}[H,a_{\omega,d}]\nonumber \\
 & =-i\omega a_{\omega,d}-\sum_{\mu}(g_{d}e^{m_{d}ik_{s}x_{\mu}})\sigma_{\mu},
\end{align}
after integrating to time, we have annihilation operator

\begin{align*}
a_{\omega,d}(t) & =a_{\omega,d}(t_{0})e^{-i\omega(t-t_{0})}\\
 & -\int_{t_{0}}^{t}ds\sum_{\mu}(g_{d}e^{m_{d}ik_{s}x_{\mu}})\sigma_{\mu}e^{-i\omega(t-s)}
\end{align*}
and

\begin{align*}
a_{\omega,d}(t) & =a_{\omega,d}(t_{f})e^{-i\omega(t-t_{f})}\\
 & -\int_{t_{f}}^{t}ds\sum_{\mu}(g_{d}e^{m_{d}ik_{s}x_{\mu}})\sigma_{\mu}e^{-i\omega(t-s)},
\end{align*}
where we set our initial condition of the time integral as initial
time $t_{0}$ and final time $t_{f}$.

Therefore, we have 

\begin{align}
a_{\omega,d}(t_{f})e^{-i\omega(t-t_{f})} & =a_{\omega,d}(t_{0})e^{-i\omega(t-t_{0})}\nonumber \\
 & +\int_{t_{0}}^{t_{f}}ds\sum_{\mu}(g_{d}e^{m_{d}ik_{s}x_{\mu}})\sigma_{\mu}e^{-i\omega(t-s)}.
\end{align}
Integrating over the frequency $\omega$, we have

\begin{align*}
\int d\omega a_{\omega,d}(t_{f})e^{-i\omega(t-t_{f})}\\
=\int d\omega a_{\omega,d}(t_{0})e^{-i\omega(t-t_{0})} & +\int d\omega\int_{t_{0}}^{t_{f}}ds\\
\sum_{\mu}(g_{d}e^{m_{d}ik_{s}x_{\mu}})\sigma_{\mu}e^{-i\omega(t-s)}.
\end{align*}
Defining the output signal operator $a_{out,d}(t)$ and the input
signal operator $a_{in,d}(t)$ as

\[
a_{out,d}(t)\equiv\frac{1}{\sqrt{2\pi}}\int d\omega a_{\omega,d}(t_{f})e^{-i\omega(t-t_{f})},
\]
 and 

\[
a_{in,d}(t)\equiv\frac{1}{\sqrt{2\pi}}\int d\omega a_{\omega,d}(t_{0})e^{-i\omega(t-t_{0})}.
\]
The input-output relation is obtained as
\[
a_{out,d}(t)=a_{in,d}(t)+\sqrt{2\pi}\sum_{\mu}(g_{d}e^{m_{d}ik_{s}x_{\mu}})\sigma_{\mu}(t).
\]
 In our system setup, there is no photonic input source for driving
atomic transitions. Therefore, $a_{in}(t)=0,$ and

\begin{equation}
a_{out,d}(t)=\sqrt{2\pi}\sum_{\mu}(g_{d}e^{m_{d}ik_{s}x_{\mu}})\sigma_{\mu}(t).
\end{equation}
With defining $\gamma_{d}\equiv2\pi g_{d}^{2}$, then output photon
flux is obtained as

\begin{equation}
\langle a_{out,d}^{\dagger}(t)a_{out,d}(t)\rangle=\langle\sum_{\mu,\nu}\gamma_{d}e^{m_{d}ik_{s}(x_{\mu}-x_{\nu})}\sigma_{\mu}^{\dagger}(t)\sigma_{\nu}(t)\rangle,
\end{equation}
where $m_{d}=+1(-1)$ for the left (right) propagation. $\gamma_{d}=\gamma_{L(R)}$
when photonic propagation direction $d$ is toward the left (right).

\bibliography{prrref}

\end{document}